# Active suppression of temperature oscillation from a pulse-tube cryocooler in a cryogen-free cryostat: Part 2. Experimental realization

Changzhao Pan[1,2, a)], Jiangfeng Hu[1,3, 4, a)], Haiyang Zhang[1,3] Yaonan Song[1,3,5], Dongxu Han[5], Wenjing Liu[1,3,5], Hui Chen[3,6], Mark Plimmer[1,2], Fernando Sparasci[1, 2], Ercang Luo[1,3,4], Bo Gao[1,3,4*], Laurent Pitre[1,2]

[1] *TIPC-LNE Joint Laboratory on Cryogenic Metrology Science and Technology, Chinese Academy of Sciences (CAS), Beijing 100190, China*

[2] *LCM-LNE-Cnam, 61 rue du Landy, 93210 La Plaine-Saint Denis, France*

[3] *Key Laboratory of Cryogenics, Technical Institute of Physics and Chemistry, Chinese Academy of Sciences, Beijing, 100190, China*

[4] *University of Chinese Academy of Sciences, Beijing 100490, China*

[5] *Beijing Institute of Petrochemical Technology, Beijing 102617, China*

[6] *Xi'an Jiaotong University, Xian 710049, China*

**Abstract:**

A cryogen-free cryostat cooled by a closed cycle cryocooler is compact, can provide uninterrupted long-term operation (up to ten thousand hours) and is suited to temperatures from 3 K to 300 K. Its intrinsic temperature oscillation, however, limits its application in experiments requiring high thermal stability at low temperature (below 77 K). Passive suppression methods are effective but all suffer from drawbacks. We describe a novel, active suppression scheme more efficient than traditional proportional-integral (PI) control. The experimental results show that it can reduce the standard deviation of the temperature oscillation by a further 30% compared with PI feedback. To the best of our knowledge, this is the first time such active suppression of temperature oscillations has been implemented with the cryogen-free cryostat. The results also show, however, that an unwanted lower frequency thermal noise will be generated, which appears to be the limit of the method. Nevertheless, the approach could be used to improve the temperature stability in all cryogen-free cryostats.

---

* Author to whom correspondence should be addressed. Email: bgao@mail.ipc.ac.cn;

a) These authors contributed to the work equally and should be regarded as co-first authors

# 1. Introduction

A cryogen-free cryostat cooled by a closed cycle cryocooler, such as a Gifford–McMahon (GM) or GM type pulse tube cryocooler (GM-PTC), can provide an uninterrupted long-term low temperature operating environment (from 3 K-300 K). Its intrinsic temperature oscillation, however, limits its application in experiments requiring high thermal stability at low temperature (especially below 77 K).Thus far, many researchers have made efforts to reduce this temperature oscillation, using the heat-capacity method [1-7], the thermal-resistance method [8-10] and the heat-switch method [11-13]. However, to the best of our knowledge, no one has yet used active suppression.

Recently, we developed a simulation model using the thermal response characteristics of a cryogen-free cryostat [14]. It revealed that PI feedback alone cannot remove the unwanted temperature modulation. In the same article, we presented a numerical simulation for a novel active stabilization scheme, the results of which were encouraging. Here we describe its experimental implementation and results obtained. The numerical values used in the theoretical companion paper [14] relate to a cryostat installed at the Technical Institute of Physics and Chemistry of the Chinese Academy of Sciences (TIPC-CAS). For practical reasons, the method was implemented experimentally at LCM-LNE-Cnam in France. The present article describes that experiment. The general results obtained are applicable to both cryostats and others elsewhere.

The remainder of the paper is structured as follows. First we present the experimental system and control method. Thereafter, the suppression results and influence factors are discussed. Finally, the current limitation and possible improvements of the method are outlined.

# 2. Apparatus and active control method

In this section we describe the cryogenic apparatus and how active suppression of the instrinsic thermal modulation was implemented.

## 2.1 Experimental system

Figure 1 shows a schematic diagram of cryostat in the experimental system. This cryostat was cooled by a commercial 4K GM-PTC (Cryomech PT405) which can supply about 0.5 W of cooling power at 4.2 K and 25 W at 65 K. To obtain a lower and more stable working temperature than that achievable with the bare cooler, the second-stage cold head and second flange were shielded by a radiation shield on the first stage. In addition, the second flange was connected to the second stage cold head using six soft thermal-links[1]. The heater (six Minco Polyimide Thermofoil heaters wired in series, total resistance about 650 Ω at room temperature) was fixed on the rods between the thermal-link and the second flange. Three calibrated negative temperature coefficient thermometers (Lakeshore Cernox 1050-CD) were used to measure the temperatures in the cryostat, namely one on the second stage cold head and two on the second flange, as shown in figure 1. In the experiment, the thermometer on the second stage cold head was used to monitor the original temperature oscillation, while that on the second flange was used to regulate and reduce it.

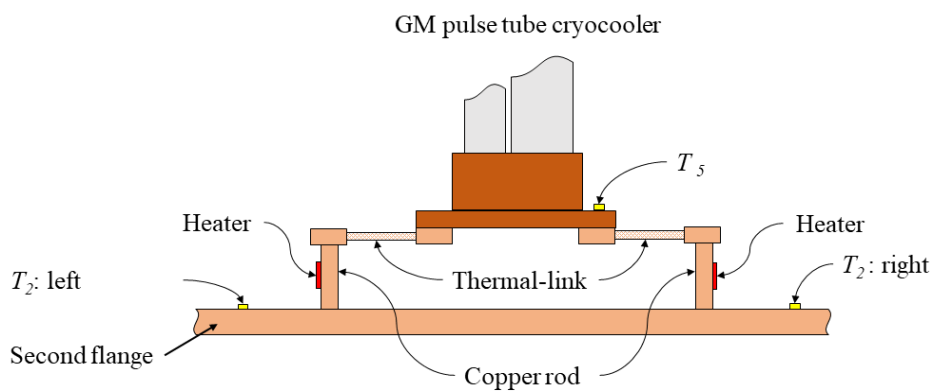

Figure 1. The schematic diagram of experimental system. Only two of the six copper rods are shown

---

[1] Material: Cu-OFE(Cu-c2) copper braid, length 62 mm, cross- sectional area 30 mm².

The resistances of all the thermometers were measured using both a digital multimeter (DMM) (Keithley 2002) and an analog-to-digital (A-to-D) interface module (National Instruments NI-9219). The DMM can measure resistance more accurately, but for 7.5 digit resolution, its minimum sampling time is a lengthy 75 ms. By contrast, the sampling time of the A-to-D interface was a mere10 ms but its 0.5 mA fixed excitation current is somewhat high, which causes more self-heating than with the DMM (current 96 μA). To benefit from the advantages of both devices, the on the second flange thermometer resistance was measured using the DMM while that of the sensor on the second stage cold head was measured via the A-to-D interface. A power supply (200 mA, 30 V) provided the heater current. Its response time is less than 10 ms over the whole range (give values), which is short enough to generate a sinusoidal heating power to balance the temperature oscillation caused by the cryocooler cycle at 1.41 Hz. A detailed comparison of results is presented in the next section.

**2.2 Implementation Strategy**

As proposed in the companion paper [14], active suppression is performed by adding an appropriate oscillating heat source to the heat generated by a proportional-integral (PI) controller. To implement the method, however, two practical problems needed to be solved. Because the oscillating temperature would either increase the value above or decrease it below its mean value in one cycle, we also needed to generate a corresponding "negative" or positive oscillating power to balance this oscillation. The first problem was thus how to avoid a negative value generated by PI feedback, which would produce no response from the heater (since a heater always supplies a positive amount of heat). The second problem was how to generate an oscillating power with a well-defined phase lag with respect to that of the temperature oscillation. Because the temperature oscillation propagates through the thermal-link, contact thermal resistance and heat capacity components, its phase undergoes some changes by the time it reaches the location of the heater. Thus, there should exist a suitable phase for balancing power oscillations.

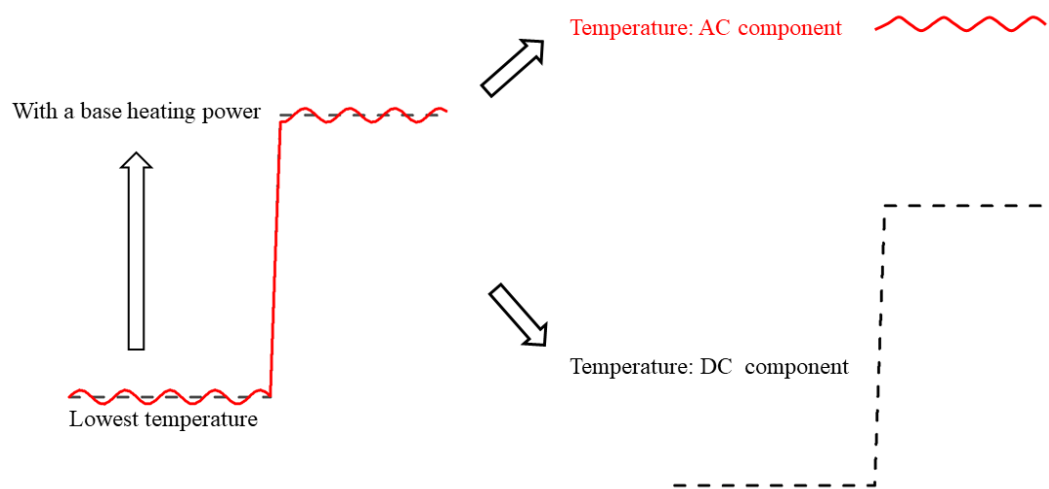

Figure 2. Strategy to produce "negative" current in the experiment. The DC component is used for PI control from one temperature to another. The AC component is used as a reference to reduce temperature oscillations.

To solve the first problem, the strategy is always to set the mean value of this oscillating heat to be greater than zero. The larger mean value set, the larger one can set the corresponding amplitude of oscillating heat. This means there is always DC heating on the flange, which raises the lowest operating temperature of the cryostat (in our case, we increased the lowest temperature by around 1.4 K). The heating generated by the heater can be split into two parts, as shown in figure 2. The DC component generated by the PI controller regulates the average temperature around the set point while the AC component provides active suppression of the temperature modulation. We first use PI feedback to regulate the temperature of the second flange from the lowest temperature (unregulated temperature around 5.6 K) up to 7 K. This heating power is relatively stable and unmodulated. The resulting temperature oscillates around the set point since, as already mentioned, PI feedback is ineffective for removing the spurious modulation. Thereafter, based on the temperature oscillation, the AC heating component is supplied using the same heater. Provided the heating power has a suitable amplitude and phase lag (see below) the temperature oscillation will be reduced.

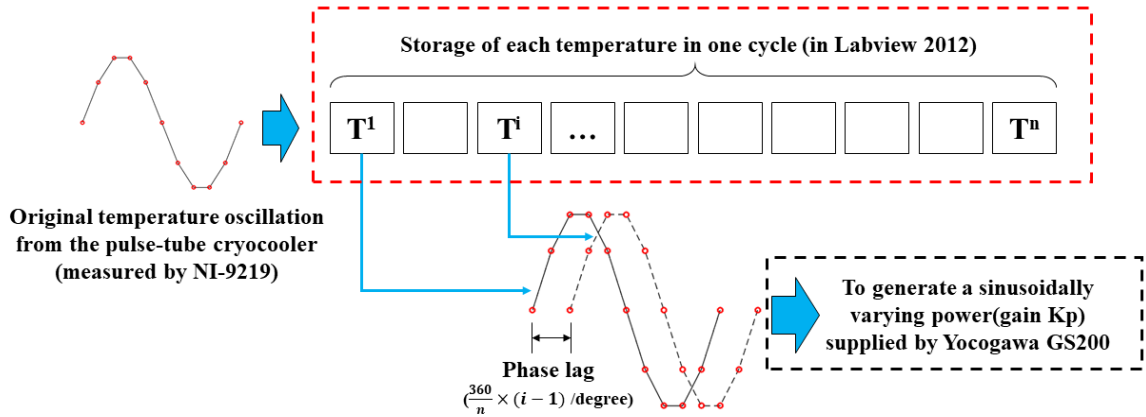

Figure 3. Experimental strategy to implement a phase-delayed heat source.

There are two ways to solve the second problem. The first is to measure the phase of temperature in real time and simultaneously synthesize a phase-shifted heating wave. This approach was tested using the function generator in LabVIEW™ software. However, it is relatively difficult to realize in practice, because it requires data be acquired and analyzed and the heat wave generated all very quickly (here in less than 0.1 s). Another, easier way is shown in figure 3. Here the temperature is measured discontinuously, $n$ times per cycle, with a uniform sampling time (80 ms in this work). The temperatures are stored as temporary variables in the LabVIEW™ software. In the experiment, we chose the $i^{th}$ storage variable as a reference to generate an oscillating heating power (we only use the $i^{th}$ temperature value to subtract its mean temperature then multiply it by a factor $Kp$). This signal is then supplied to the heater to generate a heating power with a phase lag of $(360/n) \times (i - 1)$ degrees. In the experiment, we also needed to find the optimize the phase lag (i.e. find the optimal value of $i$) and the amplification factor $Kp$.

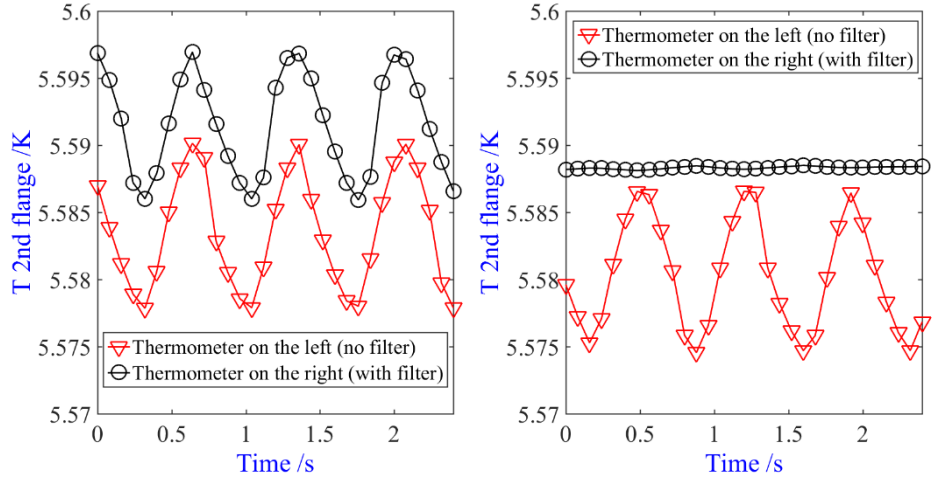

Figure 4. Measured temperature on the second flange with and without digital filtering. The solid lines are added to guide the eye

## 3. Experimental results and discussion

In this section, the factors influencing the active suppression will be discussed. These include mainly the thermometry, the phase-delay angle, the extra thermal-noise caused by the method and the different operating temperatures.

### 3.1 Temperature measurement

In this section, we describe first how the measured temperature is split into AC and DC components. Next we outline the ways to measure reference temperature and the choice of the best method. Finally, the active sinusoidal heating to suppress temperature oscillation is presented.

#### 3.1.1 DC and AC temperature components

In the experiment, two thermometers were used to measure the temperature on the second flange, one on the left, the other on the right (figure 1). The left part of figure 4 shows the temperature measurements with no digital filter. In this case, both thermometers show almost the same temperature oscillation albeit with a slightly different mean value. To separate the temperature oscillation into DC and AC components, the output of the thermometer on the right-hand side was processed with a moving digital filter (filter count: 10); the results are shown on the right side of figure 4. This signal was then used to calculate the DC heating component used for PI feedback. In addition to this DC heating, a suitable AC heating component is added and the sum of these two components is guaranteed to be a positive value, which is finally provided by the heater. At the same time, the unfiltered thermometer output is used to record and analyze the temperature oscillation.

**3.1.2 Reference temperature**

Since the oscillation heating power is generated with respect to a reference temperature, it is very important that the measured value of the latter be stable. As mentioned above, both a digital multimeter (DMM) and an A-to-D interface were be used to measure it. Figure 5 shows the outputs of the two devices (the temperature on the second stage cold head), which were measured when the temperature was regulated at 7 K (for the second flange). The upper part in figure 5 shows the results measured with the DMM (7.5-digit resolution, sampling time 80 ms). It shows some instability occurs during the measurement, which might be due to the unstable sampling rate of the DMM. If this unstable oscillation were superimposed on the reference temperature, the active sinusoidal heating would be also unstable and, while reducing the amplitude of the original oscillation, would add some other noise. The lower trace in figure 5 shows the results measured using the A-to-D interface (sampling time 40 ms, recording time 80 ms). Because of greater self-heating (excitation current 0.5 mA), the temperature is higher than the value measured by the DMM (excitation current only 96 µA). However, its measured value is relatively more stable. Because this temperature was only used as reference to generate the oscillating heating power, its absolute value was unimportant. Its stability, however, was crucial, so in later experiments, the reference temperature of the second stage was always measured using the A-to-D interface.

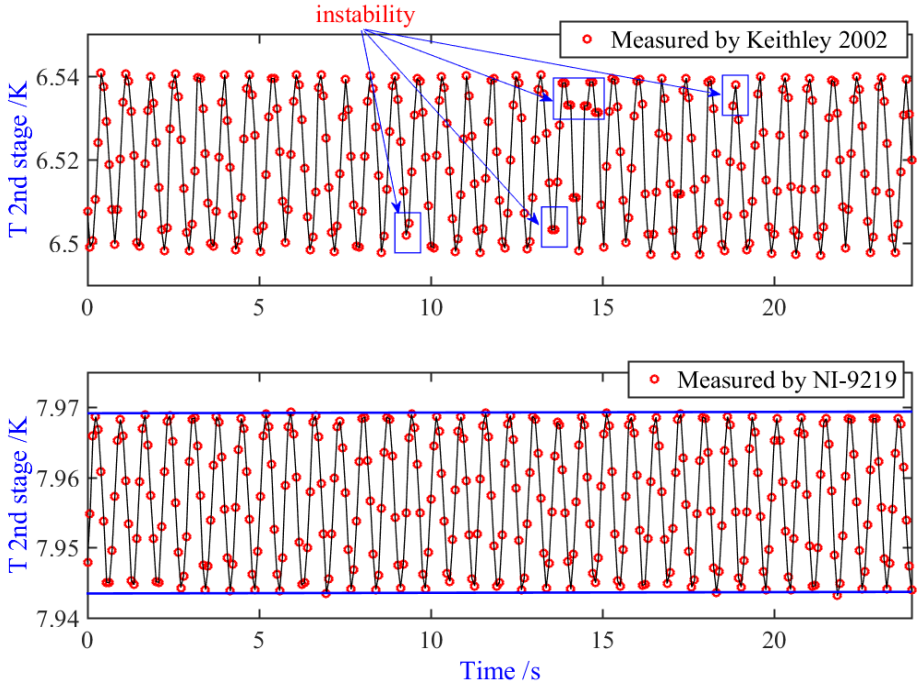

Figure 5. Stability of cryocooler second-stage temperature measured by two instruments. Upper trace: Digital multimeter (Keithley 2002) Lower trace: A-to-D interface (National Instruments NI-9219).

### 3.1.3 Active sinusoidal heating

Figure 6 shows the active oscillating heating power supplied during regulation. Its value was calculated using the reference temperature and multiplication factor $Kp = 35\,000$. Different phase delays between the oscillating heating power and the reference temperature oscillation were tested. One can see the waveform of this oscillating heating power is slightly different from that of the reference temperature. This is because the heating was the sum of DC and AC components. The DC component generated by PID controller also had some effect and changed the waveform of the final heating.

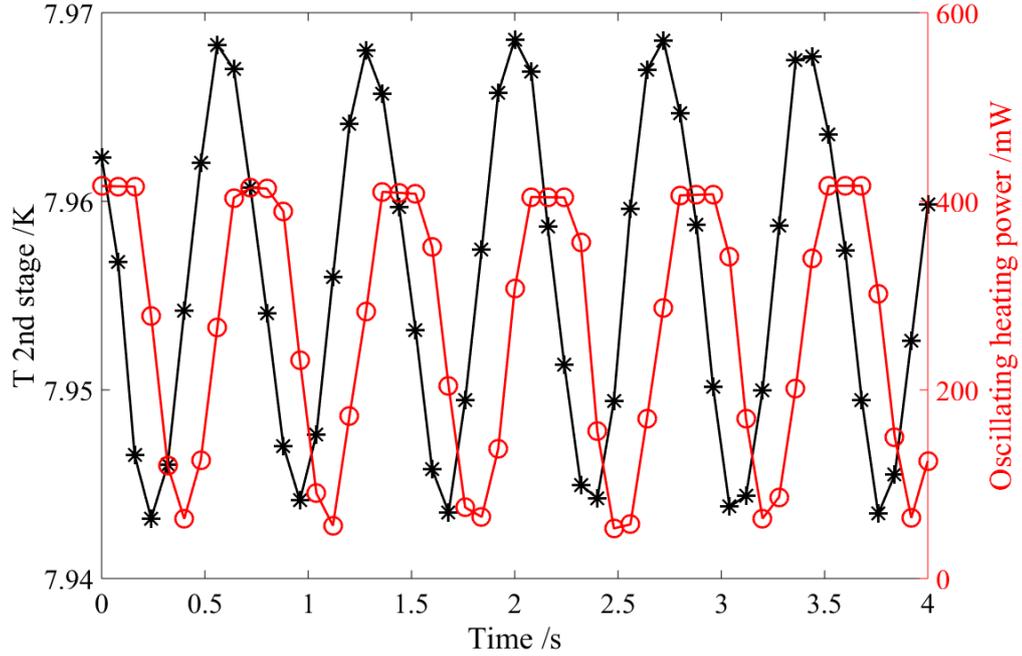

Figure 6. The oscillation heating power in the regulation. (∗) Second-stage cold head temperature, (○) Modulated heating power.

### 3.2 Optimizing thephase delay

In the experiment, the operating frequency of the GM-PTC was 1.41 Hz (the intrinsic frequency), but the sampling time in the PI controller was set to 75 ms (close to the lower limit of the DMM in 7.5-digit resolution mode), which meant the number $n$ was about 11 (figure 3). This is not accurate enough, because during the optimization process the phase lag changed by at least 36°. To increase the accuracy of optimization, $n$ was set to 21, which required the sampling time to be 35.5 ms. In the experiment, however, we did not alter the sampling time but rather calculated its value between two steps using linear interpolation. Thus, we needed to find the best phase delay among the 21 angles.

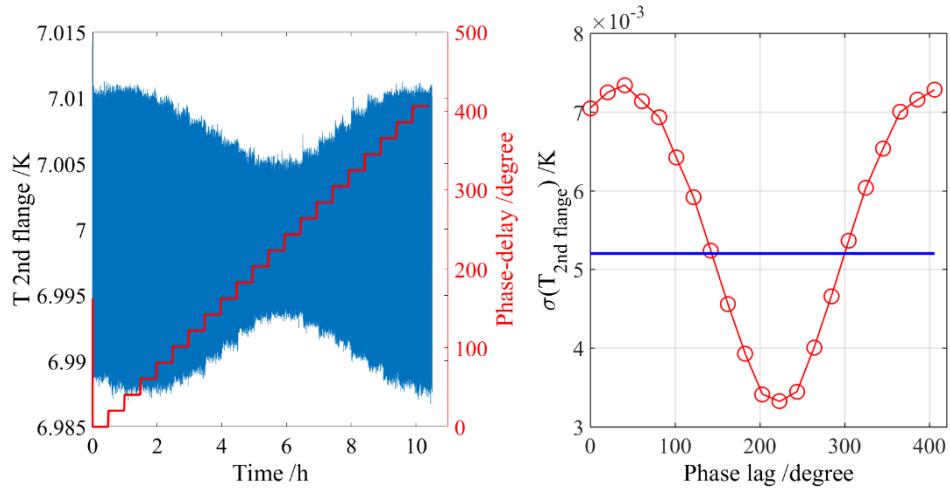

Figure 7. The effect of oscillation suppression with different phase lags. Left figure: the red step line is the phase lag (right axis), every half hour it is changed to another value. The blue line (left axis) is the temperature corresponding to each phase lag; right figure: the hollow circles (○) give the standard deviation of the temperature oscillation with different phase lags. The blue horizontal line is the standard deviation of the temperature oscillation obtained using PI control alone

To search for the optimal phase delay, an auto-scan program which could measure each phase for half an hour (21 angles from 0 ° to 360 °) was written in LabVIEW software. The left half of figure 7, shows the temperature changing during the searching process. The temperature oscillation changed significantly with different phase delay. The right half shows the standard deviation for different phase lags. The blue horizontal line is the standard deviation of temperature oscillation obtained using PI control alone, while the red circles show the standard deviation of temperature oscillation for each phase lag. One sees that, as expected, a sinusoidally modulated heating power can either suppress or amplify the original temperature oscillation. From the view of standard deviation, the optimal phase lag was about 223 °, for which the temperature oscillation was reduced from 5.2 mK to 3.3 mK.

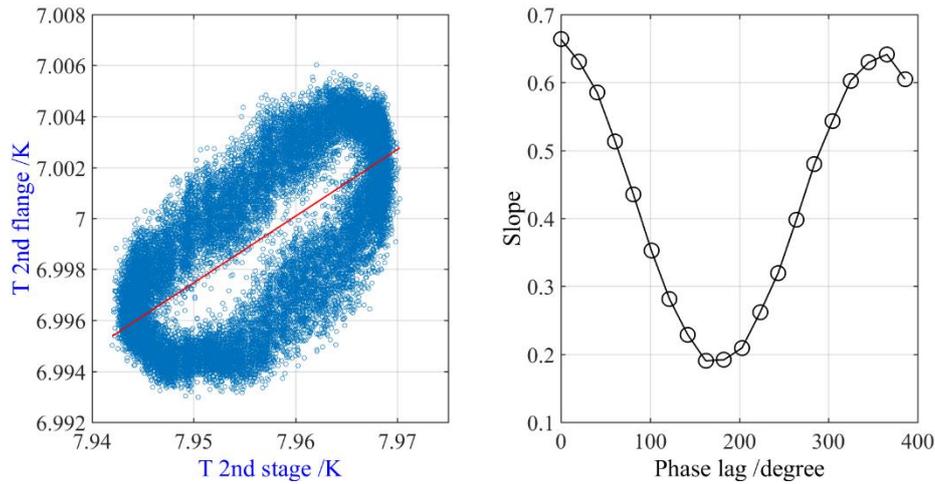

Figure 8. Left: Relationship between the temperatures on the second flange and second-stage cold end[2] (an example for phase lag of 223 °) . Right: Slopes for different phase lags

Besides using the standard deviation to evaluate the effectiveness of oscillation suppression, it can also be evaluated *via* the relationship between the temperatures on the second flange and the second stage. If the phase between the temperature on the flange and on the second stage cold head is fixed, the two temperatures should be linearly correlated, in which case all the data should lie on a straight line the slope of which could be used to quantify the efficiency of the suppression method. Figure 8 gives an example for a phase lag of 223 °, the X-axis corresponds to the temperature of second stage and the Y-axis to the temperature of the second flange. The relationship between these two temperatures was fitted to a straight line. Because the amplitude of the temperature oscillation on the second stage cold head was almost

---

[2] The temperature of the second- stage cold head is not the true value and is used only as a reference. Because of a larger self-heating, the second-stage cold head is hotter than the second flange. The x-axis in this figure shows the dynamic temperature of the second-stage cold head, while the y-axis gives the corresponding temperature of the second flange measured at the same time. If the temperatures of the second stage cold head and second flange were linearly correlated, all the data would lie on a straight line. Here, even though the correlation of those two temperatures is slightly nonlinear, for simplicity, we still use a straight line to fit them approximately and use the slope can be used to estimate the efficiency of the suppression method

unchanged (there is no change in the values along the X-axis), a smaller slope of this line corresponds to more effective oscillation suppression. The right half of figure 8 shows the slopes for different phase lags. As far as standard deviations are concerned, the smallest slope appears for a phase lag of 171 °. This difference indicates that some other changes occur during the process of oscillation suppression, which will be discussed in next section.

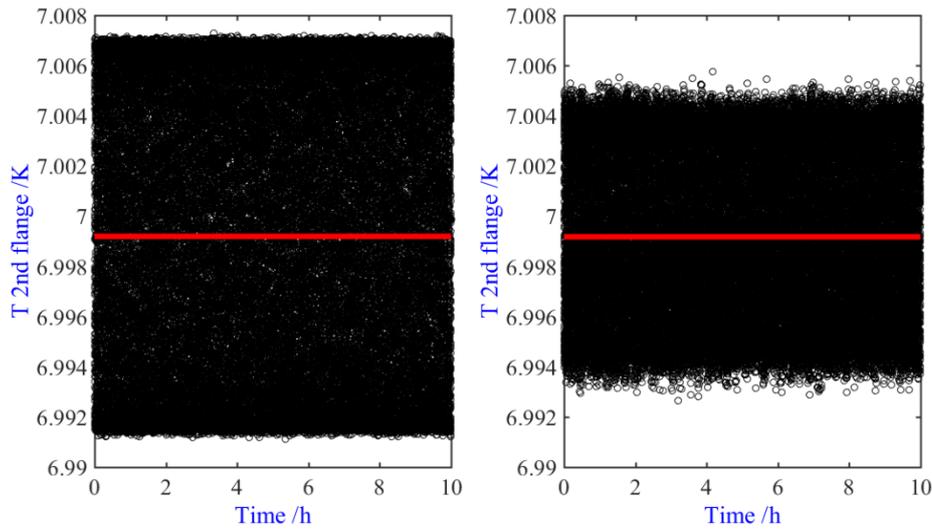

Figure 9. Long-term stability of active oscillation suppression. The horizontal lines indicate the mean values. Left: only using traditional PI control (standard deviation: 5.2mK); right: using active control with phase-delay 223° (standard deviation: 3.2 mK)

After optimization of the phase delay, a long-term measurement was performed to check temperature stability. Figure 9 shows the results for 10 hours of continuous thermometry. The left-hand side shows the results obtained using traditional PI control and the right-hand side those using active control. There is a visible improvement, which highlights the effectiveness of the latter method.

## 3.3 Associated thermal-noise

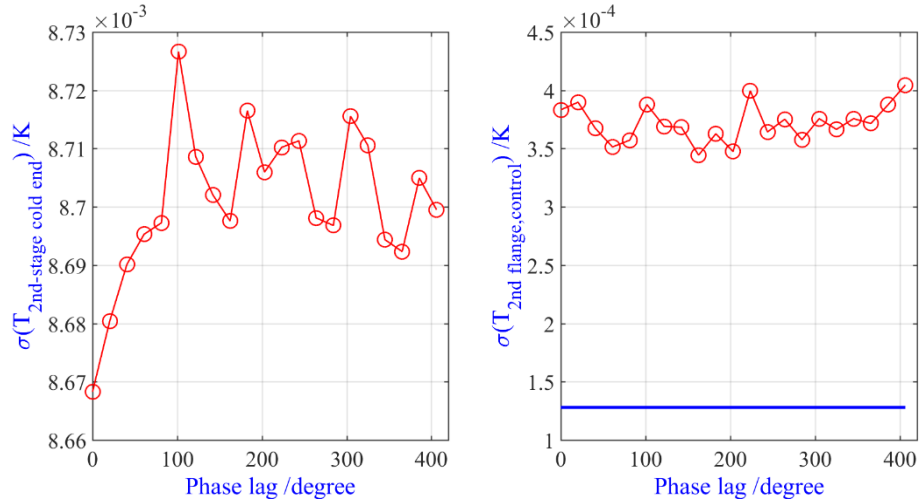

Figure 10. The standard deviation of temperature on the second stage and the standard deviation of DC component temperature on the second flange for different phase lags.

Why is the optimal phase lag for temperature stabilization not that which yields the smallest slope? Figure10 shows the standard deviation of two more temperatures obtained using different phase lags. The graph on the left shows the standard deviation of the temperature of the second stage cold head while the one on the right shows the standard deviation of the temperature (DC component) of the second flange. The horizontal line in the right-hand graph corresponds to the standard deviation before active oscillation suppression. Neither temperature has a clear relationship with the phase lag. For the DC component temperature on the second flange, however, owing to thermal noise added by active suppression, its standard deviation is always higher by about 0.3 mK. This extra thermal noise slightly changes the slope of the fit in figure 8.

To understand this additional thermal noise more deeply, consider the transient temperature oscillation before and after active oscillation suppression shown in figure 11. It is clear that the oscillation suppression reduces the fluctuation amplitude of the temperature wave but also makes its mean temperature unstable. When the amplification factor *Kp* is increased, the suppression effect increases (smaller oscillation amplitude) but the frequency of temperature oscillation is modified (*i.e.* it falls). Figure 12 shows a Fourier analysis of temperature oscillation before and after the active oscillation suppression, revealing the appearance of a lower frequency temperature wave in the latter case. This low-frequency thermal noise adds extra instability to the system, thereby hindering the effectiveness of active suppression. As discussed above, this thermal noise was mainly caused by the instability of the mean temperature (DC component). Thus, improving the stability of the reference temperature measurement and PID controller might reduce it.

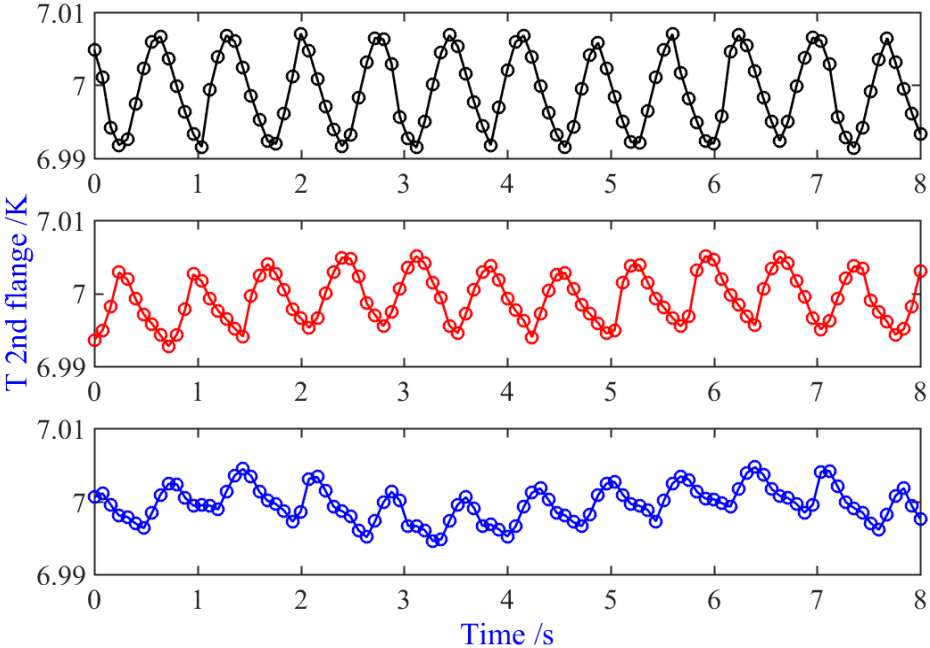

Figure 11 Transient temperature oscillation with and without active suppression for different values of PI parameters. upper: PI feedback ($Kp = 20\,000$, $I = 0.6$) without active suppression, middle: PI feedback ($Kp = 20\,000$, $I = 0.6$) plus active suppression ($Kp = 35\,000$), lower: PI feedback ($Kp = 20\,000$, $I = 0.6$) plus active suppression ($Kp = 50\,000$)

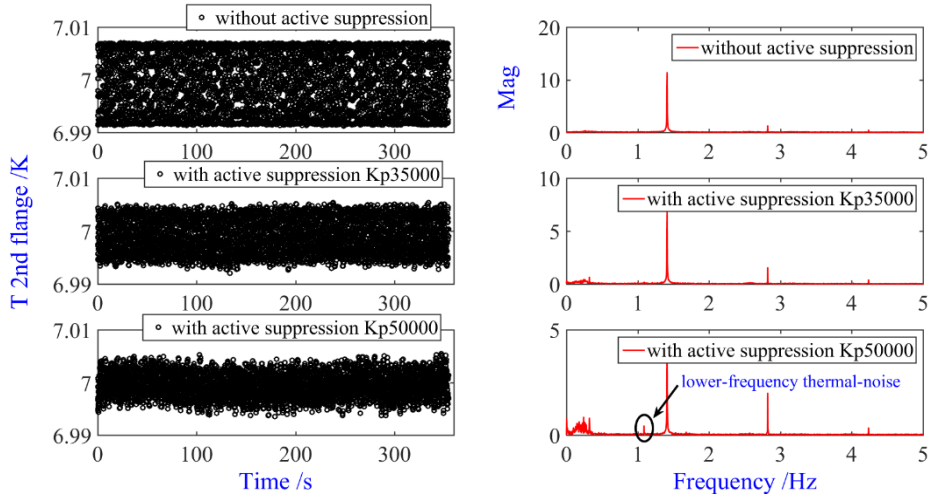

Figure 12 Fourier analysis (right) of the temperature oscillation (left) before and after active suppression. The attenuation of the fundamental oscillation at 1.4 Hz appears to add extra ones, a small component near 1.1 Hz, one at the second harmonic as well as a broad peak from 0 to 0.2 Hz

### 3.4 Results for a higher temperature (9 K)

To demonstrate the active suppression method is applicable at temperatures other than 7 K, it was also tested at a higher temperature, namely 9 K. Here temperature oscillations, though a little smaller than at 7 K[3], still subsist and need to be counterbalanced. Figure 13 shows the corresponding experimental results. Three different phase delay angles were measured for different amplification factors $Kp$. One sees this active method attenuates the temperature oscillation significantly (the standard deviation is reduced by 1.5 mK). One also sees that the optimal phase lag is about 223 ° and that the suppression improves with increasing values of the proportional gain $Kp$.

---

[3] The heat-capacity of the material used for the thermal link (copper) increases with temperature, which helps to damp the temperature oscillation from the cryocooler. At a higher temperature, therefore, the amplitude of temperature oscillation should be smaller.

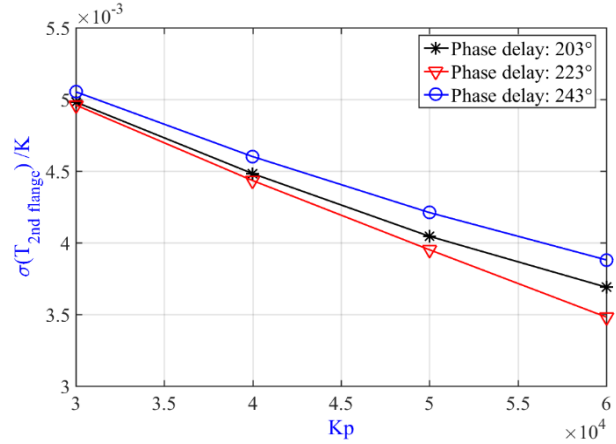

Figure 13. The effect of oscillation suppression at 9 K for three different phase lags

## 4. Conclusion

A new active control method to suppress the temperature oscillation from a 4K GM pulse tube cryocooler has been tested experimentally. It involves the active generation of an oscillating heat source to compensate the periodically varying temperature. To implement the method, a reference temperature was stored and used to generate a sinusoidally varying heat source with a well-defined phase lag. The results showed there exists an optimal phase lag for which the standard deviation of the oscillation is reduced by 30% compared with that obtained using traditional PI control. At the same time, however, increasing the proportional gain of the oscillating heating power generates additional, lower frequency thermal-noise (in addition to a second harmonic component) which ultimately limits the effectiveness of the method. This limitation could be reduced by using a more stable reference temperature measurement and PI controller. Nevertheless, we have demonstrated the feasibility of active oscillation suppression method and recommend it be used to improve the temperature stability in all cryogen-free cryostats.

**Acknowledgments**

This work was supported by the National Key R&D Program of China (Grant No. 2016YFE0204200), the National Natural Science Foundation of China (Grant No. 51627809), the International Partnership Program of the Chinese Academy of Sciences (Grant No. 1A1111KYSB20160017) and the European Metrology Research Programme (EMRP) project18SIB02 "Real-K". Changzhao Pan was supported by funding provided by a H2020 Marie Skłodowska Curie Individual Fellowship -2018 (834024). The authors also wish to thank Nicola Chiodo and Pascal Gambette for their help in setting up the apparatus.